\RequirePackage{fix-cm}
\documentclass[twocolumn,natbib]{svjour3}    		      % twocolumn

\usepackage[utf8]{inputenc}
\usepackage[T1]{fontenc}
\usepackage{lmodern}

\usepackage{siunitx,upgreek}
\usepackage{graphicx}
\usepackage{amsmath,amssymb}
\usepackage{xspace}							% For spaces in newcommand
\usepackage{color}
\usepackage{multirow}

\usepackage{hyperref}
\hypersetup{
	colorlinks=true,
	allcolors=blue,
}

\graphicspath{{./figs/}}
\usepackage{txfonts}				% For times font. Must load after ansmath
\usepackage[misc]{ifsym}			% For envelope signe \Letter

\renewcommand{\deg}[0]{^\circ} 

\providecommand{\e}[1]{\ensuremath{\times 10^{#1}}}
\newcommand{\pd}[2]{\frac{\partial #1}{\partial #2}}
\newcommand{\pdd}[2]{\frac{\partial^2 #1}{\partial #2^2}} 

\renewcommand{\d}[2]{\frac{\text{d} #1}{\text{d} #2}} 
\newcommand{\ol}[1]{\overline{#1}}
\renewcommand{\vec}[1]{\boldsymbol{#1}} 
\newcommand*{\chem}[1]{\ensuremath{\mathrm{#1}}}
\newcommand{\me}{\mathrm{e}}
\newcommand{\el}{\mathrm{e^-}}

\newcommand{\Sh}{\mbox{\textit{Sh}}}  % Sherwood number
\newcommand{\Sc}{\mbox{\textit{Sc}}}  % Schmidt number
\newcommand{\Sr}{\mbox{\textit{Sr}}}  % Strouhal number
\newcommand{\Pen}{\mbox{\textit{Pe}}}  % Peclet number
\newcommand{\Rey}{\mbox{\textit{Re}}}  % Peclet number

\newcommand{\ainv}{\alpha_{\text{inv}}}

\newcommand{\aq}{\alpha_{\text{q}}}

\newcommand{\sq}{s_{\text{q}}}

\newcommand{\Sinv}{S_{\text{inv}}}
\newcommand{\ssob}{s_{\text{sob}}}
\newcommand{\Ssob}{S_{\text{sob}}}
\newcommand{\Sana}{S_\text{ana}}

\newcommand{\she}{\Sh_{\text{exp}}}

\newcommand{\shes}{\Sh^*_{\text{exp}}}
\newcommand{\klev}{k_\text{q}}
\newcommand{\kcot}{k_{\text{cot}}}
\newcommand{\uaxe}{u_\text{cl}}

\newcommand{\lev}{L\'{e}v\^{e}que\xspace}
\newcommand{\sob}{Sobol\'ik\xspace}

\newcommand{\LADYF}{{\fontfamily{lmr}\selectfont\textup{\normalsize L\kern -.35em\lower -.5ex\hbox{\scriptsize A} \normalsize\kern -.4em\lower .2ex
		\hbox{\footnotesize D}\kern -.24em\lower .ex\hbox{Y}\kern -.1em
		\hbox{\scriptsize F}}}}

\begin{document}
% ==================== TITLE AUTHOR ===================== %
\title{Two-component instantaneous wall shear rate measurements in reversing flows}
\author{M.-{\'E}. Lamarche-Gagnon \and
	V. Sobol{\'i}k \and
	J. V\'etel
}

\institute{
	M.-{\'E}. Lamarche-Gagnon (\,\Letter\,) \and J. V\'etel \at
		\LADYF, Department of mechanical engineering \\
		Polytechnique Montr\'eal, Canada \\
		m-e.lamarche-gagnon@polymtl.ca; jerome.vetel@polymtl.ca
	\and
	V. Sobol{\'i}k \at
		LaSIE, Faculty of Science and Technology\\
		Universit\'e de La Rochelle, France\\
		vsobolik@univ-lr.fr
}
\date{Received: date / Accepted: date}

\maketitle

% ====================== ABSTRACT ======================= %
\begin{abstract}
%\sloppy
Validation experiments of the two-dimensional inverse algorithm proposed by \citet{lama2017inv} are performed in a pulsed Poiseuille flow exposing shear reversal phases. The method is applied to the three-segment electrodiffusion (ED) probe for which a specific nondimensionalization process is suggested, allowing to better link measurements from a real ED probe to the modeled one in the inverse problem. This approach provided a two-com\-po\-nent wall shear rate in good agreement with the one obtained from laser Doppler anemometry (LDA) measurements, thus validating the ability of ED probes to deal with high-amplitude unsteady flows. The classic linear velocity approximation ($u=sy$) in the probe vicinity is also investigated in such a flow.
\end{abstract}

%% ==================================================== %%
%% ==================================================== %%
\section{Introduction}\label{sec:intro}
Measurement of wall shear stress in unsteady flows remains a challenge despite the many techniques developed in the last century \citep{naug2002mode}. When both its magnitude and direction are sought, when space and time resolutions are an issue, the number of reliable methods falls dramatically. Hot-film and electrodiffusion (ED) probes both allowed many authors to perform near-such measurements in turbulent flows \citep{mitc1966stud,alfr1988fluc,sirk1970limi,he2011wall,desl2004near}, but none truly handled all the above concerns simultaneously. On the one hand, heat conduction to the wall considerably affects the hot-film unsteady response while, on the other hand, both hot-film and ED probes are not known to accurately assess the wall shear stress direction in unsteady flows, especially when dealing with large amplitude fluctuations.

Some methods introduced in recent years have the potential to overcome these difficulties in determining the magnitude and direction of instantaneous wall friction, in particular the ones developed by \citet{gros2006} and by \citet{liu2014}. The former captures the deflection of elastomeric micro-cylinders embedded in a wall using high-speed cameras, while the latter evaluates the mass transfer caused by the sublimation of a paint overlaid on a surface. Although both of these methods are promising, their implementation may be rather complex and their scope is currently limited.

\begin{figure*}
	\centering
	\includegraphics[scale=0.9]{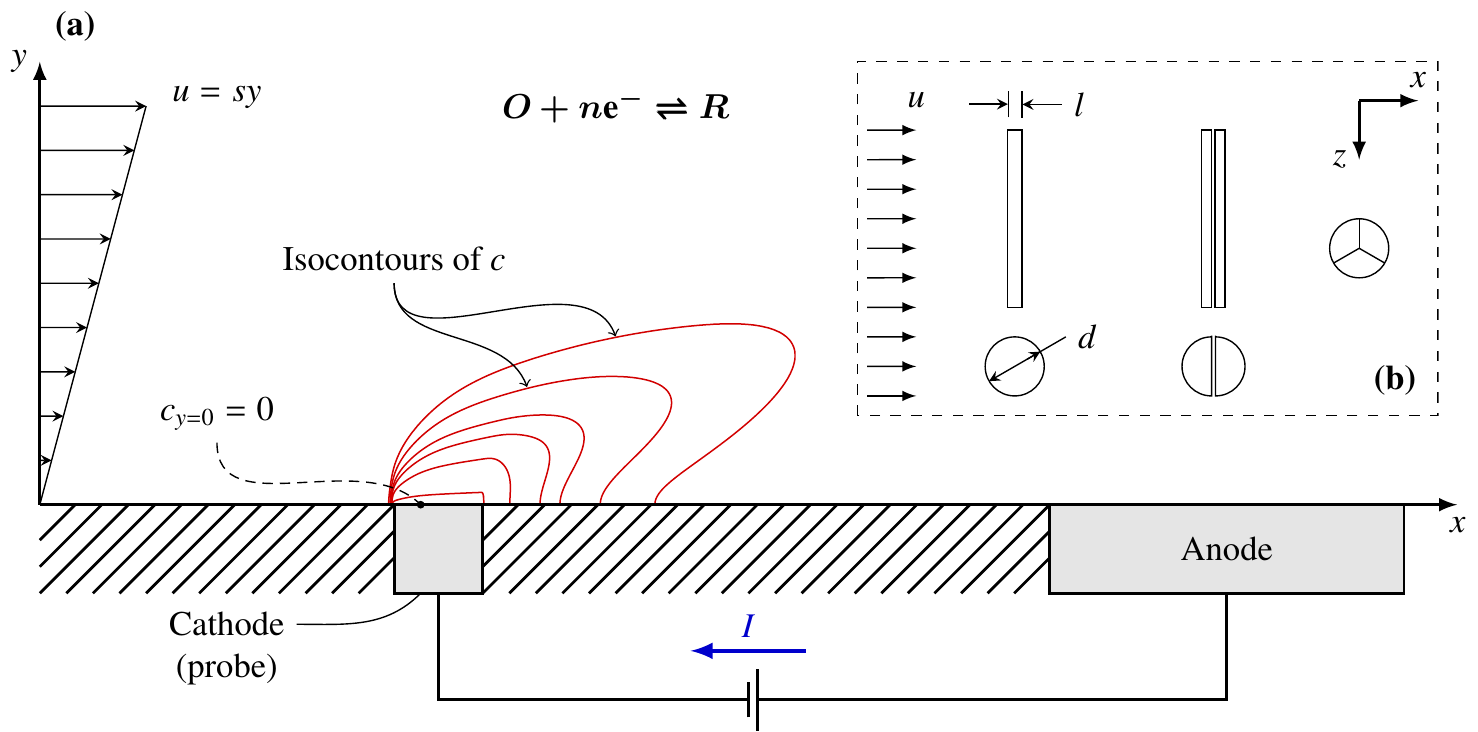}
	\caption{\textbf{a} Schematic of the ED method basis. \textbf{b} Types of probes, viewed from above. From left to right: single (rectangular and circular), double (or `sandwich') and three-segment probes.}
	\label{fig:01}
\end{figure*}

\cite{lama2017inv} recently proposed a post-treatment method to deal with the so-called capacitive effects on each individual segment's signal of a three-segment ED probe, allowing one to obtain the proper instantaneous two-component wall shear rate in unsteady flows. The approach is based on an adaptation of the inverse method \citep{mao1991analysis,rehimi2006inverse} to the two-dimensional wall shear rate. While the method was numerically validated by \cite{lama2017inv}, the main purpose of the present article is to demonstrate its ability to deal with a real three-segment probe signal and procure accurate measurements under periodic high-amplitude fluctuations of the wall shear rate, even in the case of shear reversal. In the following sections, the basis of the ED method and inverse algorithm will first be recalled. Then, the experimental apparatus and instrumentation will be introduced in Sect.~\ref{sec:expres}, followed with an investigation of the flow inside the test section and a description of the proposed nondimensionalization procedure. Section~\ref{sec:results} presents the inverse method results for different experimental cases along with complementary wall shear rate measurements performed with laser Doppler anemometry (LDA). The linear velocity approximation, which is a fundamental principle in the ED technique, is also examined in cases involving flow reversal.

%% ==================================================== %%
\subsection{Principles of the ED method and main post-treatments}
The ED method relies on the mass-transfer between a redox couple ($O$--$R$) contained in the solution and electrodes flush-mounted to a wall (Fig.~\ref{fig:01}). When a sufficiently large voltage is imposed between the anode and the cathode, the reacting species at the probe surface is completely depleted ($c_{y=0}=0$), resulting in local strong concentration gradients. Under such conditions, the electrochemical reaction rate is at its maximum and the so-called \textit{limiting current} $I$ flows through the electrodes and solution \citep{selman1978mass}. As the Schmidt number $\Sc=\nu/D$ is very large in most ED cells ($\Sc\sim1000$), where $\nu$ is the kinematic viscosity and $D$ the diffusion coefficient, one can expect the hydrodynamic boundary layer to be very thin when compared to the diffusion layer so as to assume a linear velocity profile $u=sy$ in the probe vicinity, with $s$ the wall shear rate magnitude. When convection effects are dominant in a steady flow, the \lev or \textit{quasi-steady} solution states that $s$ is proportional to the cube of the current \citep{reiss1963expe}, namely
\begin{equation}
   s_q = (I/\klev)^3,
   \label{eq:Slev}
\end{equation}
with $\klev=0.80755nFAC_0l^{-1/3}D^{2/3}$ and where $n$, $F$, $A$, $C_0$, $l$ respectively refer to the number of electrons involved in the reaction, the Faraday constant, the sensor area, the concentration in the bulk solution and the probe size, although constant $\klev$ is also accessible through an appropriate calibration. When the flow is unsteady, the most common and accessible post-treatment, commonly known as the \sob method \citep{sobolik1987simultaneous}, takes advantage of the signal time derivative to deal with the capacitive effects in the diffusion layer. The instantaneous wall shear rate then follows
\begin{equation}
    \ssob = \sq + \frac{2}{3}\chi\sq^{-2/3}\d{\sq}{t},
\end{equation}
with $\chi=0.80755^{-2}\pi^{-1}l^{2/3}D^{-1/3}$ or, using the information from the Cottrell asymptote calibration \citep{sobolik1998calibration}, can be written as $\chi=(\kcot/\klev)^2$.

%% ==================================================== %%
\subsection{Statement of the problem}
\sob method offers a straightforward correction and is fairly accurate even when unsteadiness is dominant, but fails in the presence of shear reversal. Moreover, when one also seeks the 360\,$\deg$ shear direction using, for instance, three-segment probes (cf. Fig.~\ref{fig:01}b), only the quasi-steady approach proposed by \cite{wein1987theory} is available, valid for low-frequency processes. The two-component inverse algorithm \citep{lama2017inv} was developed especially to deal with high-amplitude unsteady flows where, from a single three-segment probe signal, both instantaneous wall shear rate magnitude $s$ and direction $\alpha$ can be assessed. The aim of the present paper is to experimentally validate the inverse algorithm and further explore its limitations when treating real signals. Tests are performed in a high-shear rectangular channel, where the flow is controlled so as to generate periodic fluctuations on both $s$ and $\alpha$ over a broad frequency range. The small channel size (height $\sim3\,$mm) allows one to produce a flow featuring a large time-averaged wall shear stress while preserving a laminar flow, which was essential to procure accurate reference values (cf. Sect.~\ref{sec:Sref}) for calibration and validation purposes.

%% ==================================================== %%
\subsection{Inverse algorithm}
Under normal ED conditions\footnote{i.e. when a background electrolyte is added in excess in the solution so as to counter migration of the $O$--$R$ ions.}, mass transfer measured by an ED probe is essentially controlled by diffusion effects, which are manifested by concentration gradients in the probe vicinity (cf. Fig.~\ref{fig:01}). The overall process is then governed by the convection--diffusion (CD) equation:
\begin{equation}
	\pd{c}{t} + \vec{u}\cdot\nabla c = D\nabla^2 c
	\label{eq:CD}
\end{equation}
where $\vec{u}=sy\left(\sin\alpha\vec{i} + \cos\alpha\vec{k}\right)$ and $\alpha$ is the wall shear rate direction; in its dimensionless form, one obtains
\begin{multline}
	\Sr\pd{C}{\tau} +SY\left(\sin\alpha\pd{C}{X} + \cos\alpha\pd{C}{Z}\right) = \\ 
	\Pen^{-2/3}\left( \pdd{C}{X} + \pdd{C}{Z}\right) + \pdd{C}{Y},
	\label{eq:cdadim3D}
\end{multline}
using the following dimensionless variables
\begin{gather}
	\begin{array}{c}
	X = \dfrac{x}{d},  \qquad 
	Y = \dfrac{y}{d}\Pen^{1/3},  \qquad  Z = \dfrac{z}{d},  \qquad  \tau = tf,  \qquad  	\\ \vspace{-5pt}\\ 
	\Pen = \dfrac{\ol{s}d^2}{D},  \qquad 
	\Sr = \dfrac{fd^{2/3}}{\ol{s}^{2/3}D^{1/3}},  \qquad  S = \dfrac{s}{\ol{s}},  \qquad  C = \dfrac{c}{C_0},
	\end{array}
	\label{eq:adim}
\end{gather}
where $d$ is the (equivalent) diameter of a circular (three-segment) probe, $f$ is a characteristic frequency and $\Sr$, $\Pen$ are the Strouhal and P\'eclet numbers, respectively. $\ol{(\sim)}$ indicates a time-averaged quantity over one period.

\begin{figure}
	\centering
	\includegraphics[scale=0.78]{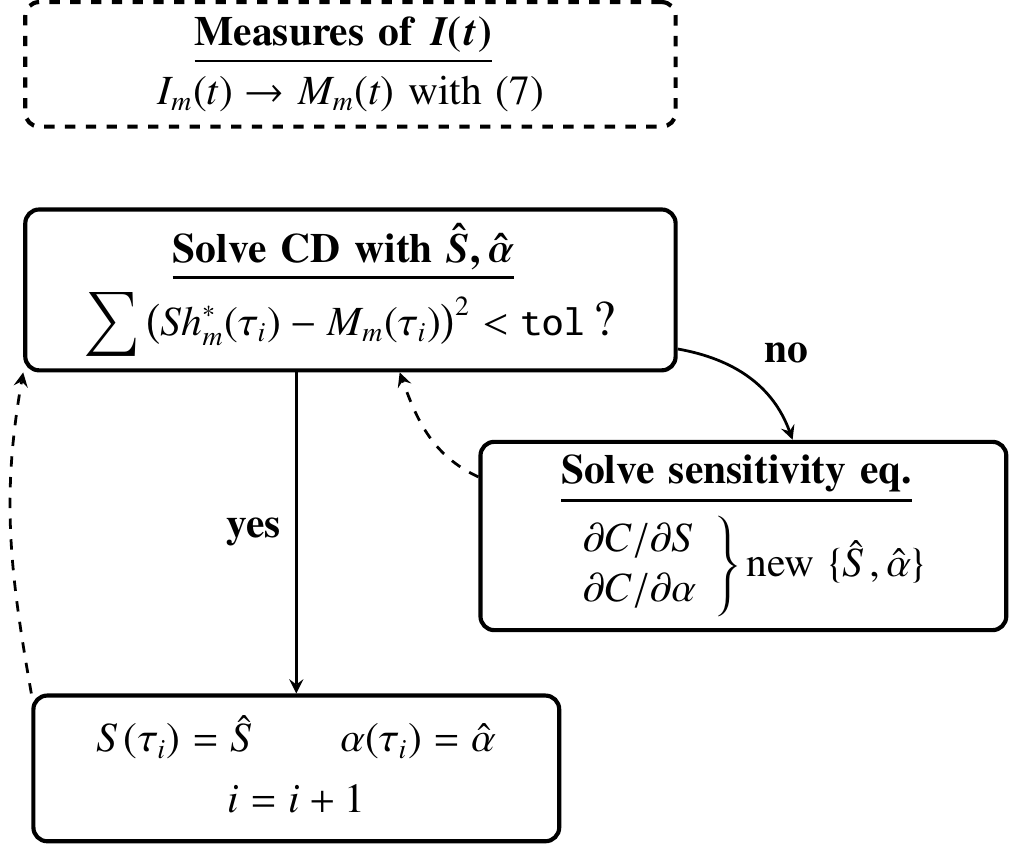}
	\caption{Main steps of the inverse algorithm. $M_m=\Sh^*_{\text{exp},m}$ is the measured modified Sherwood number from segment $m\in\{0,1,2\}$. Superscript $\hat{(\sim)}$ stands for a guessed value. Procedure is repeated for all time steps $i$ until a tolerance \texttt{tol} is reached.}
	\label{fig:02}
\end{figure}

The \textit{inverse problem} consists in iteratively solving the \textit{direct problem}, i.e. the CD equation \eqref{eq:cdadim3D} at given $\Pen$ and $\Sr$, by adjusting the input parameters $S$ and $\alpha$ until the numerical results converge to the measurements, here represented by the Sherwood number of segment $m\in\{0,1,2\}$
\begin{equation}
	\Sh_{\text{exp},\,m} = \frac{I_md}{nFADC_0}
	\label{eq:shexp}
\end{equation}
or by the \textit{modified} mass transfer coefficient
\begin{equation}
	\Sh^*_{\text{exp},\,m}=\Sh_{\text{exp},\,m}\Pen^{-1/3},
\end{equation}
while its numerical counterpart is evaluated using
\begin{equation}
	\Sh^*_{\text{num},\,m} = \frac{1}{A}\iint_{A_m}\left.\pd{C}{Y}\right|_{Y=0}dA,
	\label{eq:shnum}
\end{equation}
with $A_m$ the area of the discretized segment $m$ while $A$ stands for the total area. Sensitivity equations associated to \eqref{eq:cdadim3D}, namely equations for $\partial C/\partial S$ and $\partial C/\partial\alpha$, are simultaneously solved in the algorithm to evaluate the appropriate corrections $[dS,d\alpha]$ at each time step. Figure \ref{fig:02} summarizes the proposed procedure. Note that for the remainder of the paper, the notation $M=\shes$ and $\Sh^*=\Sh^*_\text{num}$ will be used for convenience unless otherwise specified. The reader is referred to the work of \cite{lama2017inv} and \cite{rehimi2006inverse} for more details. 

\section{Methodology \label{sec:expres}}
\begin{figure*}
	\centering
	\includegraphics[scale=1.1]{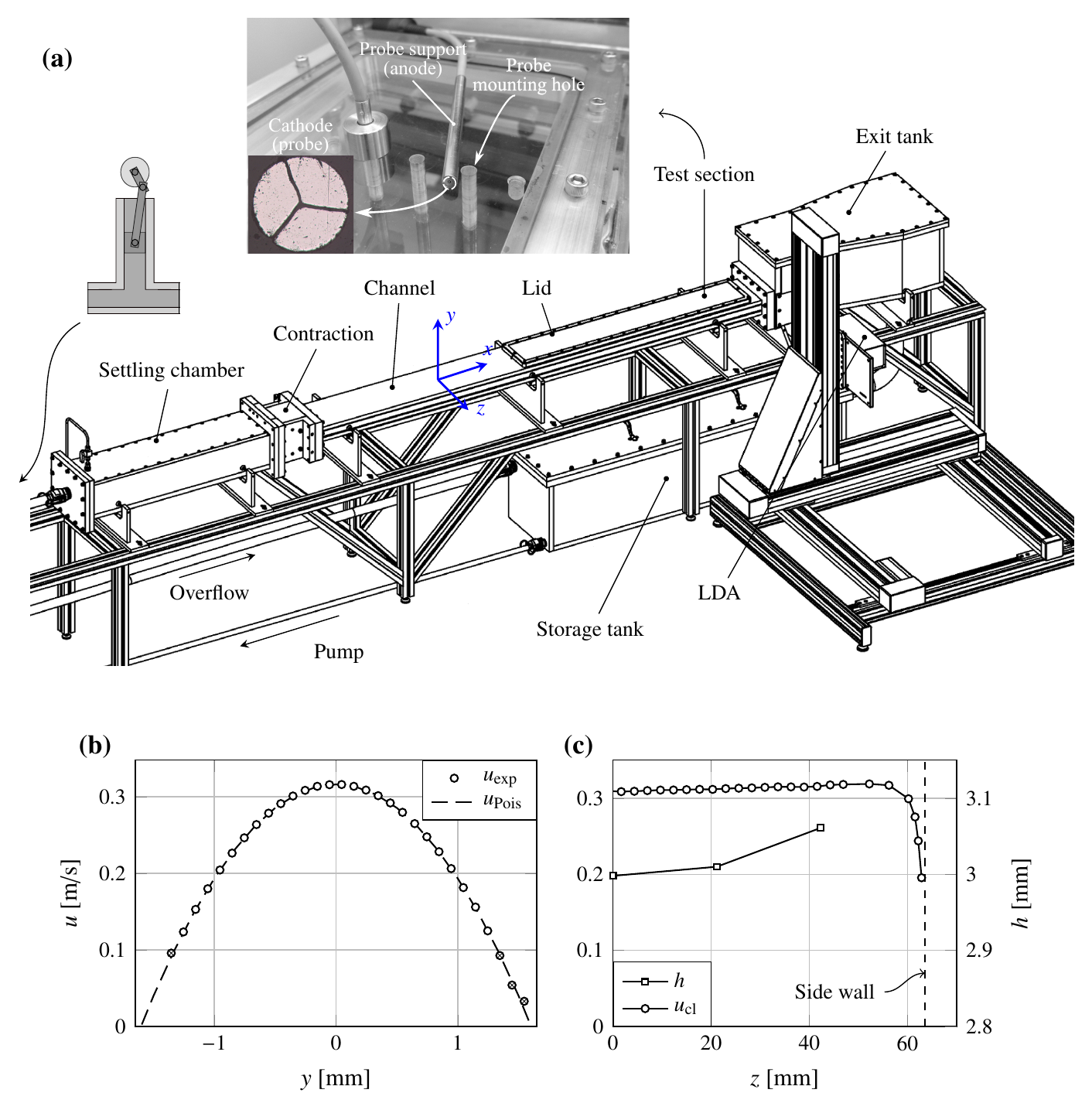}
	\caption{\textbf{a} Experimental apparatus. The photograph exposes the top of the test section with two ED probes. \textbf{b} Velocity profile $u_\text{exp}(y)$ at a position $z=W/6$, where the transverse coordinate $|z|<W/2$ and $W=127\,$mm is the inside width of the channel. Also shown is the Poiseuille profile resulting from a least-squares fit between \eqref{eq:u0} and measurements $u_\text{exp}$. $\otimes$ marks refer to the closest wall positions and were not used in the regression due to the greater measurement error. \textbf{c} Transverse velocity profile $\uaxe(z)$ (at $y=0$) and channel height measurements at the three probe's locations (see top of \textbf{a}).}
	\label{fig:3}
\end{figure*}

\subsection{Setup and instrumentation}
The small-scale water tunnel (Fig.~\ref{fig:3}) of the fluid dynamic laboratory of Polytechnique Montr\'eal (\LADYF) was used for the mass transfer measurements. Fluid is first pumped to an elevated tank subdivided into sections so as to ensure a constant fluid head and isolate the main flow from the pump perturbations. The test section is part of a 2\,m long rectangular channel of cross section $127\times3.1$\,mm with an aspect ratio $\sim40$. Velocity fluctuations in the test section are reduced thanks to a large settling chamber including a set of four screens with decreasing mesh size and one honeycomb. This is followed by a smooth two-dimensional contraction (fifth order polynomial curved walls with an area contraction ratio $\sim32$). Overflow tubes in both the elevated and exit tanks ensure a constant fluid head, while the flow rate can be carefully adjusted using a low pressure drop electronic globe valve. A circulating bath coupled to a heat exchanger provides temperature control in the whole setup. Fully developed near two-dimensional Poiseuille flow is achieved at the electrodiffusion probes' location, positioned near the exit of the channel. Flow is maintained laminar for Reynolds numbers $\Rey=4h\uaxe/3\nu$ up to $\sim3500$, with $h$, $\uaxe$ the channel height and the centerline axial velocity ($y=0$), respectively. A piston pump driven by a stepper motor controls the unsteady motion of the flow; in particular, periodic high-amplitude fluctuations can be generated up to a frequency of $\sim 20\,$Hz. Shear reversal in the test section was achievable for $f\lesssim8$\,Hz. 

\begin{figure*}
	\centering
	\includegraphics{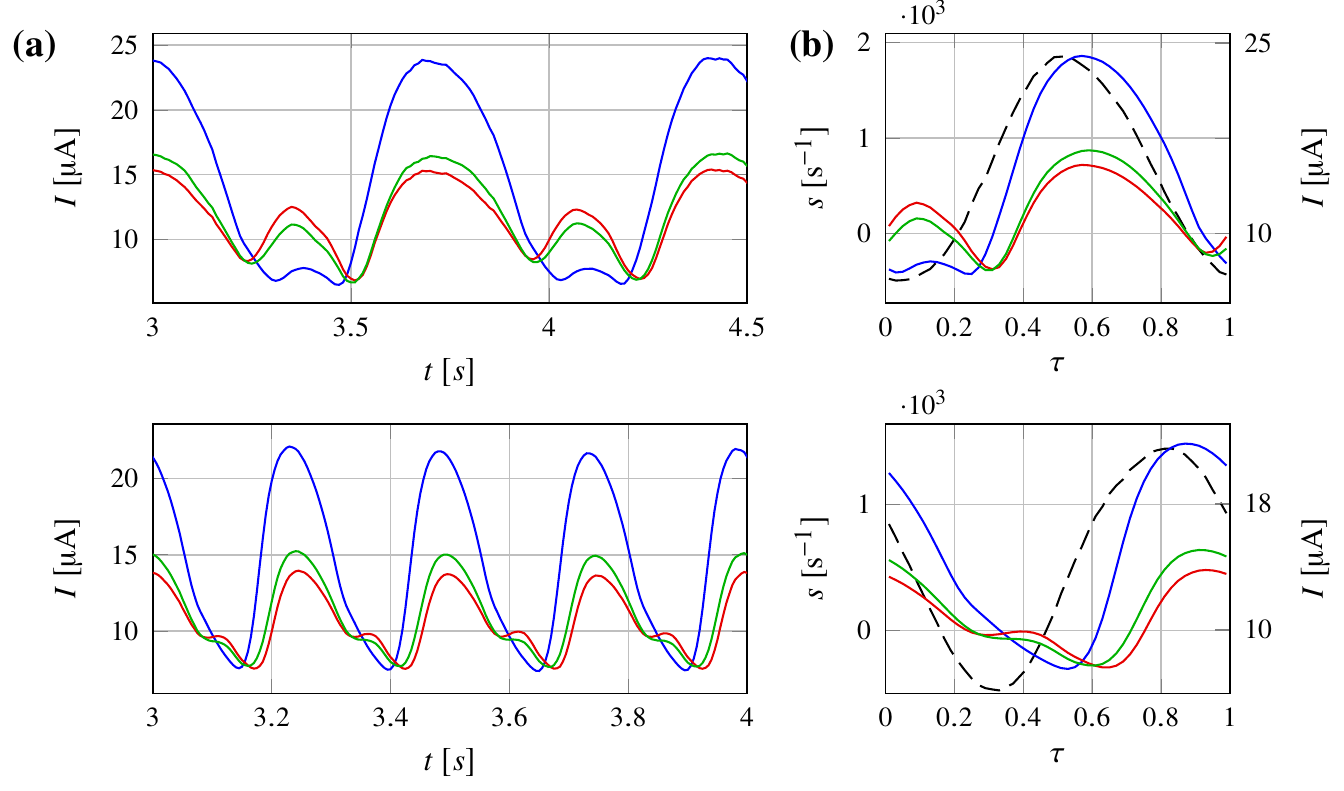}
	\caption{\textbf{a} Typical three-segment probe raw signals, each curve being the current of a different segment. A low sampling frequency $f_s$ (top: case 0, $f_s=100$\,Hz; bottom: case 1, $f_s=200$\,Hz; see Table \ref{tab:parexp}) combined with a high quality ADC ensured very clean high SNR signals. \textbf{b} Phase-averaged $I_m(t)$ and wall shear rate (dashed lines) signals.}
	\label{fig:4}
\end{figure*}

The provided three-segment probe was assembled according to the method of \citet{sobolik1991three}. The working segments made of platinum are glued into an epoxy resin and mounted on a stainless steel rod, which also serves as the anode. The diameter of an equivalent circular probe having the same effective area as the three segments is close to 0.52\,mm and was established using an optical microscope (see top of Fig.~\ref{fig:3}a); this value is $\sim10$ times smaller than the diameter of the stainless steel rod. Probes were first polished with fine emery paper, then using alumina slurries of decreasing particle size down to 0.05\,$\upmu$m. Routine polishing was executed using the finest slurry and, more frequently, by simply rubbing the probe surface with a soft cloth. Circular electrodes of three different diameters were also used for some experiments; all cathodes are mounted on the same type of rods (Fig.~\ref{fig:3}a) and so the probes' location could be easily varied between each other. Sensors are positioned in the second half of the channel, referred to as the test section. This segment is characterized by a removable lid allowing control on the probes mounting, in particular to ensure that they are well flush-mounted. Considering the small channel size ($h\sim3$\,mm), great care is crucial when positioning the probes as a vertical displacement of $\sim0.1$\,mm could lead to substantial errors on the probe signal. The redox couple ferricyanide--ferrocyanide was used as depolarizer using a bulk concentration $C_0=0.025$\,M (equimolar) in water. Reduction of ferricyanide at the working electrode (cathode) was considered, as per reaction
\begin{align*}
\chem{[Fe(CN)_6]^{3-}} + \el \longrightarrow \chem{[Fe(CN)_6]^{4-}}.
\end{align*}
This was achieved under a potential of $-0.6\,$V, corresponding to a central position in the limiting current plateau. 0.15\,M of $\chem{K_2SO_4}$ was added to counter migration effects; such concentrations of the species, providing an overall fluid density $\rho=1.03$\,g/cm$^3$, was also chosen to match the LDA seeding particles density and hence produce a uniform particles distribution. Dynamic viscosity measured at 22.0$\,\deg$C was $\mu=1.00\e{-3}$\,Pa\,s. Diffusion coefficient obtained by chronoamperometry at this temperature was $D=7.5\e{-10}\,$m\,s$^{-2}$, which is very close to common tabulated values \citep{hanr1996meas,bard2001electrochemical}. Mass transfer signals are amplified through a current follower and then recorded using a 24 bits low-noise ADC; sampling rate was varied from 100\,Hz to 500\,Hz depending on cases. Typical raw three-segment probe signals are illustrated in Fig.~\ref{fig:4}.

The water tunnel was made especially for ED studies. Hence, all elements in contact with the fluid are made of chemical resistant plastic material such as PVC or PMMA. A laser Doppler anemometry (LDA) apparatus mounted on a 3D traverse was used for flow characterization, probe calibration and to provide a reference value for the unsteady wall shear rate (see Sect.~\ref{sec:Sref}). The channel and test section are thus exclusively made of PMMA for LDA access from the side walls. As cast PMMA flatness on such long section is poor, upper and lower walls were first flatten and polished to a transparent finish, ensuring a surface roughness close to cast PMMA ($R_\text{a}=0.011$\,mm). Note that the same properties hold for the test section as the whole process was conducted with the lid fastened to ensure a smooth transition. Although a finite gap is unavoidable at the test section inception because of the lid, LDA measurements in this region showed no significant flow perturbations. 

All signals were synchronized by triggering the mass transfer probes recording using the LDA apparatus.

\subsection{Wall shear rate reference value \label{sec:Sref}}
Wall shear rate in steady two-dimensional laminar Poiseuille flow is expressed by 
\begin{equation}
	s = \frac{4\uaxe}{h}.
	\label{eq:Smean}
\end{equation}
In regions where the flow is fully developed, $s$ is thus available with a single LDA measure instead of, for instance, evaluating the velocity gradient very close to the wall. This latter method is here very imprecise considering the thickness of the LDA measurement volume ($\sim0.1$\,mm) compare to $h$, where strong velocity gradients are expected inside the volume itself; LDA errors in close wall regions are thus amplified (as observed in Fig.~\ref{fig:3}b). The accessible distance from the wall is also limited by laser reflections. Since the velocity profiles $u(y)$ measured in the test section perfectly matches the theoretical Poiseuille equation (Fig.~\ref{fig:3}b), the former method for evaluating $s$ was assumed to be accurate. Quadratic regression on such a velocity profile was used to precisely position the laser on the channel centerline. Considering that errors associated with LDA velocity measurements at the centerline are small (low RMS and velocity gradients), the accuracy on $s$ is thus limited by the precision on $h$, evaluated to $\sim 5$\% or less. As per \eqref{eq:Smean}, this error proportionally affects $s$. Despite the carefulness in the fabrication process, upper and lower walls of the channel appear to be slightly curved in the transverse $z$ direction, likely to be caused by the polishing procedure. This results in a slight variation of the channel height $h$ between the side walls (located at $z=\pm W/2$) and the center ($z=0$), with $h$ being larger near the side walls; this variation is estimated to be lower than $0.1\,$mm (see Fig.~\ref{fig:3}c). While counter-intuitive at first, the augmentation of $h$ goes with an acceleration of the centerline velocity $\uaxe(z)$ as shown in Fig~\ref{fig:3}c, which effect has been confirmed by numerical simulations using a similar curved wall channel. Moreover, simulations confirm that the mean flow can be considered to be locally two-dimensional, namely that \eqref{eq:Smean} holds using the local height $h(z)$.

\paragraph{Fluctuating wall shear rate: }the analytical solution of the oscillating Poiseuille flow with null mean flow (over one period, $\ol{u}=0$) is derivable~\citep{schl2000}; in particular, for the developed flow between two parallel plates, the $x$ momentum equation is reduced to
\begin{align}
\pd{u}{t} = -\frac{1}{\rho}\pd{p}{x} + \nu\pdd{u}{y},
\label{eq:momen}
\end{align}
where a pressure gradient of the form
\begin{align}
-\frac{1}{\rho}\pd{p}{x} = \sum_{n=1}^{N}K_n\sin(\omega_n t)
\label{eq:gradP}
\end{align}
is imposed, with $K_n$ a constant associated to a periodic solicitation of frequency $f_n$ and $\omega_n=2\pi f_n$. The pressure gradient is then defined as a combination of $N$ harmonic oscillations of amplitude $K_n$. Using complex notation and supposing that the velocity profile $u_n(y,t)$, associated with frequency $f_n$, is of the form 
$$u_n(y,t)=g(y)\me^{i\omega_n t},$$ 
\eqref{eq:momen} for velocity $u_n$ becomes
\begin{align}
g'' - g\frac{i\omega_n}{\nu} &= \frac{iK_n}{\nu}.
\label{eq:momcomp}
\end{align}
Solving for $g$, the contribution $u_n(y,t)$ is obtained: 
\begin{equation}
u_n(y,t) = - \frac{K_n}{\omega_n}\me^{i\omega_n t} \left[1-\frac{\cosh\left(y\sqrt{i\omega_n/\nu}\right)}{\cosh\left(h/2\sqrt{i\omega_n/\nu}\right)}\right].
\label{eq:un}
\end{equation}
By virtue of the superposition principle (\eqref{eq:momen} is linear), one can obtain the complete unsteady velocity profile by summing the contributions:
\begin{equation}
u(y,t) =\sum_{n=1}^{N}u_n(y,t).
\label{eq:untot}
\end{equation}
From \eqref{eq:un} and \eqref{eq:untot}, the fluctuating wall shear rate is then given by
\begin{align}
s(t) &= \left.\pd{u}{y}\right|_{y=\pm h/2} \nonumber \\ 
&= \pm \sum_{n=1}^{N} \frac{K_n}{\omega_n}\me^{i\omega_nt} \sqrt{i\omega_n/\nu} \; \tanh\left(h/2\sqrt{i\omega_n/\nu}\right). 
\label{eq:Sn}
\end{align}
Equations \eqref{eq:un} and \eqref{eq:Sn} hence suggest that $s(t)$ can be assessed from the time trace of the velocity $u(y,t)$ at a single $y$ location over one period, considering that $K_n$ are known constant values. In particular, a centerline velocity measurement is appealing, where \eqref{eq:un} becomes
\begin{align} 
u(0,t) = - \sum_{n=1}^{N} \frac{K_n}{\omega_n}\me^{i\omega_nt} \left[1-\frac{1}{\cosh\left(h/2\sqrt{i\omega_n/\nu}\right)}\right]. 
\label{eq:uncentre}
\end{align}
The flow being periodic, measures of $\uaxe(t) = u(0,t)$ can be decomposed in Fourier series. Using \eqref{eq:uncentre} along with the Fourier coefficients so-obtained, the $K_n$ constants can be determined and then used in \eqref{eq:Sn} for $s(t)$ evaluation. Although this has been developed considering a two-dimensional flow, the 3D effects caused by the walls curvature described earlier only has little effect on $u$ considering the small variation of $h(z)$ and the $\cosh$ value in \eqref{eq:uncentre}. This has also been verified numerically and experimentally, where the unsteady $\uaxe(t)$ is rather constant in the transverse direction away from the side walls.

\begin{figure*}
	\centering
	\includegraphics{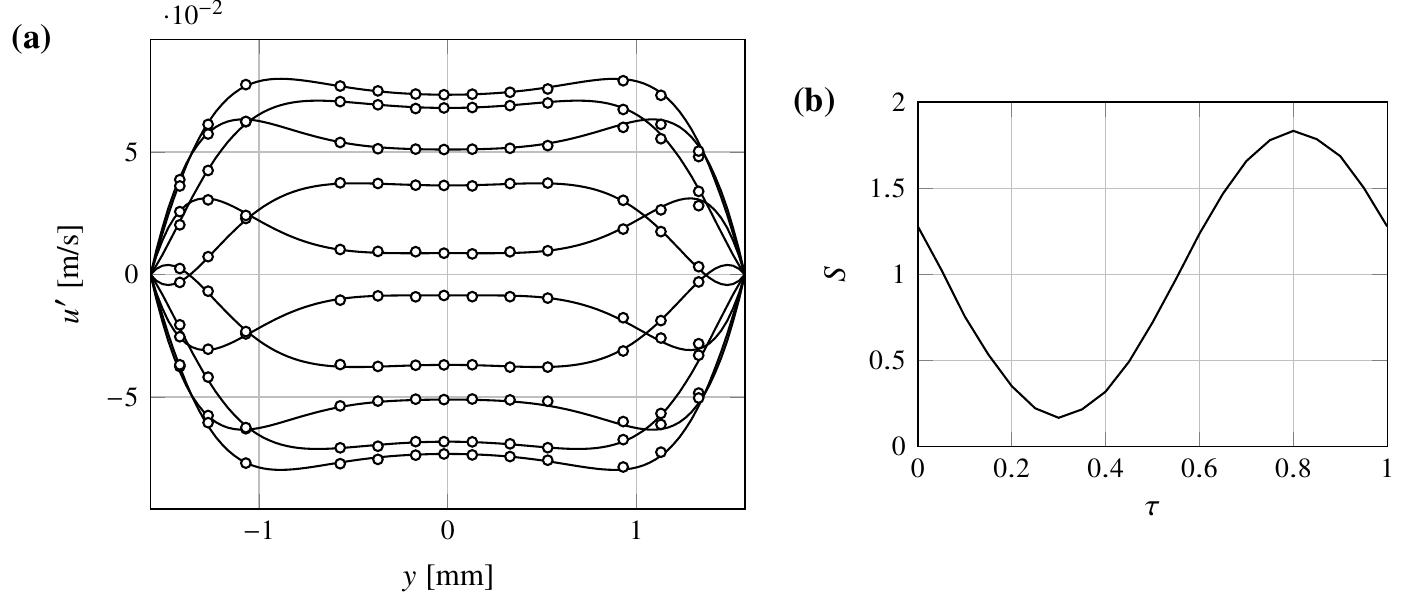}
	\caption{\textbf{a} Fluctuating velocity with height inside the test section of the periodically pulsated channel flow ($f=3$\,Hz, $\ol{\uaxe}=0.32$\,m/s, $\Rey=1375$). Analytical solution \eqref{eq:untot} is represented by solid lines and is plotted at different instants in the period. Circular marks correspond to phase-averaged LDA instantaneous measurements at those same instants; measures were subtracted with the time-averaged value $\ol{u}(y)$ at the corresponding height $y$. \textbf{b} Corresponding dimensionless wall shear rate $S$ plotted over one period of oscillation, calculated with \eqref{eq:Sn} and normalized with its time-averaged value. Note that a similar concordance between $\eqref{eq:untot}$ and LDA measurements is also observed for cases exposing shear reversal.}
	\label{fig:14b}
\end{figure*}

When a non-zero time-averaged pressure gradient is added to \eqref{eq:gradP}, one could expect the velocity profile to be the sum of an unsteady part $u'(y,t)$ as per \eqref{eq:untot} and the steady Poiseuille profile, namely
\begin{align}
u(y,t) = \ol{u}(y) + u'(y,t)
\label{eq:ucomplet}
\end{align}
with
\begin{align}
\ol{u}(y) = \ol{\uaxe}\left[1-\left(\frac{2y}{h}\right)^2\right].
\label{eq:u0}
\end{align}
Equation \eqref{eq:ucomplet} has been verified experimentally with LDA measurements of unsteady velocity profiles at different pulsation frequencies. Measures were phase-averaged to reduce random noise in the signals. As LDA sampling is irregular and essentially a function of the fluid velocity, more periods were required for positions closer to the walls, where   
\begin{enumerate}
	\item fluid velocity is lower, reducing the sampling rate and
	\item velocity gradients are stronger, enhancing the measurement error and intensity of random noise.
\end{enumerate}
Between 100 and 1000 periods were used depending on the signal quality and frequency for a proper convergence of averages. Strong concordance is observed between measurements and the analytical model given by \eqref{eq:untot}\footnote{\eqref{eq:untot} instead of \eqref{eq:ucomplet} is compared to better visualize the differences with the measurements. A similar concordance is also noted for the complete profile \eqref{eq:ucomplet}.} for positions near the channel centerline ($y=0$), while more disparities are observed closer to the wall (see Fig.~\ref{fig:14b}a). This is to be expected since the centerline position is used as reference in the Fourier decomposition. Nevertheless, an offset as low as 0.02\,mm on the $y$ positions close to the wall in \eqref{eq:un} and \eqref{eq:u0} would produce a perfect fit with the analytic model. While such error could be attributed to the traverse mechanism, the exact cause for theses discrepancies remains unknown. Considering the overall measurement errors, we shall consider that \eqref{eq:ucomplet} adequately represents the experimental flow.

With \eqref{eq:ucomplet} validated, one could further expect that the unsteady wall shear rate be the combined effect of the time-averaged $\ol{s}$ of \eqref{eq:Smean} superposed with an unsteady $s'$ from \eqref{eq:Sn}, namely
\begin{align}
\left.s(t)\right|_{y=\pm h/2} = \mp\frac{4\ol{\uaxe}}{h} 
\pm \sum_{n=1}^{N} \frac{\lambda_nK_n}{\omega_n}\me^{i\omega_nt}  \tanh\left(\lambda_nh/2\right). 
%	s(t) &= \left.\pd{u}{y}\right|_{y=\pm h/2}\nonumber\\
%	&= \ol{s} + s'(t)
\label{eq:Stot}
\end{align}
with $\lambda_n=\sqrt{i\omega_n/\nu}$. Once again, this extension is based on a two-dimensional flow and experimental validation would corroborate its usage considering the curvature and side wall effects. Yet, a precise direct measurement of the instantaneous wall shear rate is hardly achievable, especially due to the channel size. \eqref{eq:Stot} was then not validated experimentally, but numerical simulations for the oscillating laminar channel flow were again performed. Results exposed a strong concordance with the analytical model \eqref{eq:Stot}, where $s(t)$ was compared at different transverse positions using the local height $h(z)$ in both \eqref{eq:Smean} and \eqref{eq:Sn}. Actually, as earlier exposed for the centerline fluctuating velocity $\uaxe'(t)$, $s'(t)$ is almost invariant in the transverse direction owing to such minor variations of $h(z)$. This cannot be concluded for $\ol{s}$ whose value slightly increases toward the side walls ($\sim1\,\%$ between, for example, positions $z=W/3$ and $z=0$); such variations in the transverse direction is less than that of $h$ itself considering $\ol{u}$ also increases with $h$ (see \eqref{eq:Smean} and Fig.~\ref{fig:3}c).

In spite of the preceding discussions, \eqref{eq:Stot} will thus be considered as the \textit{true} wall shear rate; an error of $\sim5\,\%$ is estimated on the mean wall shear rate $\ol{s}$, arising especially from the value of $h$. A typical $S(\tau)$ evolution is plotted in Fig.~\ref{fig:14b}b.

\subsection{Nondimensionalization of the current $I$ \label{sec:calnum}}
Recalling \eqref{eq:shexp}, a simple or direct nondimensionalization method uses experimental data alone, namely 
\begin{equation}
	\she = \frac{Id}{nFAC_0D},
	\label{eq:Shdef2}
\end{equation}
which can essentially be written as $\she=KI$ with $K$ a constant. However, this procedure is likely to lead to erroneous results or, at the very least, cause offsets on $S$ and $\alpha$ during the inverse process because of the unavoidable errors in \eqref{eq:Shdef2} parameters. First, $d$ and $A$ values are often deduced by optical means to obtain the so-called \textit{geometric area} $A_\text{geo}$ and its equivalent diameter for a perfectly circular probe, which does not account for potential inactive parts of the probes caused by probe poisoning \citep{selman1978mass} or adsorbed species on its surface. Such effects are hardly quantifiable and tend to modify the \textit{effective area} so-obtained. Moreover, it is of general opinion \citep{arvia67,hanr1996meas} that the diffusion coefficient should not vary with the concentration of the constitutive species in the solution. Yet, we noted that the measured value for the ferricyanide diffusivity can vary up to 20\,\% among authors, even when adjusting the temperature and viscosity effects using the Einstein-Stokes extension $D\propto T/\mu$. Methods like the rotating disk electrode or chronoamperometry (Cottrell asymptote) are common and accessible methods for measuring $D$, giving access to the \textit{effective diffusivity} considering migration effects (sometimes non-negli\-gi\-ble) and other phenomena that could occur within the diffusion layer \citep{selman1978mass}. While this could partly explain those discrepancies, diffusivity measurement is actually quite fastidious, mainly because the sensor area $A$ is met in most methods and adsorption effects are especially important in those experiments; errors on $A$ are thus reflected on $D$. Furthermore, ferrocyanide and, to a lesser extent, ferricyanide are known to deteriorate when exposed to light \citep{selman1978mass,berger1983opt}. Some authors detected a degradation of the ferricyanide concentration $C_0$ as much as 10\,\% following the first 2--3 days of the solution preparation, even when the solution was kept in a darkened room \citep{szanto2008limiting}; later on, the degradation is much slower and the concentration tends to stabilize. On the present authors' experience, however, no such significant variations were observed when comparing for example values of the calibration coefficient $\klev$ (see \eqref{eq:Slev}) days even weeks following the preparation. Variations were observed indeed, but without an actual trend, suggesting that those were more likely caused by surface alteration of the probe from routine polishing processes or probe displacement during manipulations. This reinforces the hypothesis that deterioration of $C_0$ would mainly happen short after the solution preparation. 

%While inactive sectors do not participate to the reaction and thus reduce the \textit{effective area}, the area actually tends to the geometric one in extended electrolysis like steady processes. On the other hand, some adsorbed species can enhance the reaction \citep{bard2001electrochemical}. In extended electrolysis like steady processes, the \textit{effective} area tend to the geometric one as the diffusion layer thickening can cover those imperfections. In unsteady processes
\begin{figure}
	\centering
	\includegraphics{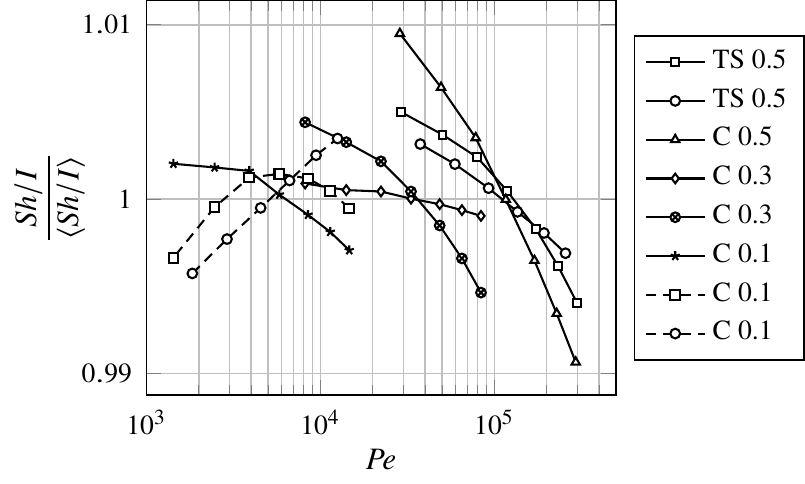}
	\caption{Ratio between numerical Sherwood number $\Sh$ and experimental current $I$ at various $\Pen$, normalized with the average value of the curve $\left<\Sh/I\right>$. Each mark corresponds to a different flow condition ($\ol{s}$) while curves and symbols refer to distinct probes. $I$ measurements were made using six circular (C) and two three-segment (TS) ED probes; $\Pen$ is calculated with LDA measurements using $\ol{s}$ from \eqref{eq:Smean} and with the corresponding diameter of the sensor (approximated values in the legend are in millimeters). $\Sh$ is obtained by solving the stationary direct problem at the corresponding $\Pen$.}
	\label{fig:15}
\end{figure}

\begin{figure*}
	\centering
	\includegraphics[scale=0.8]{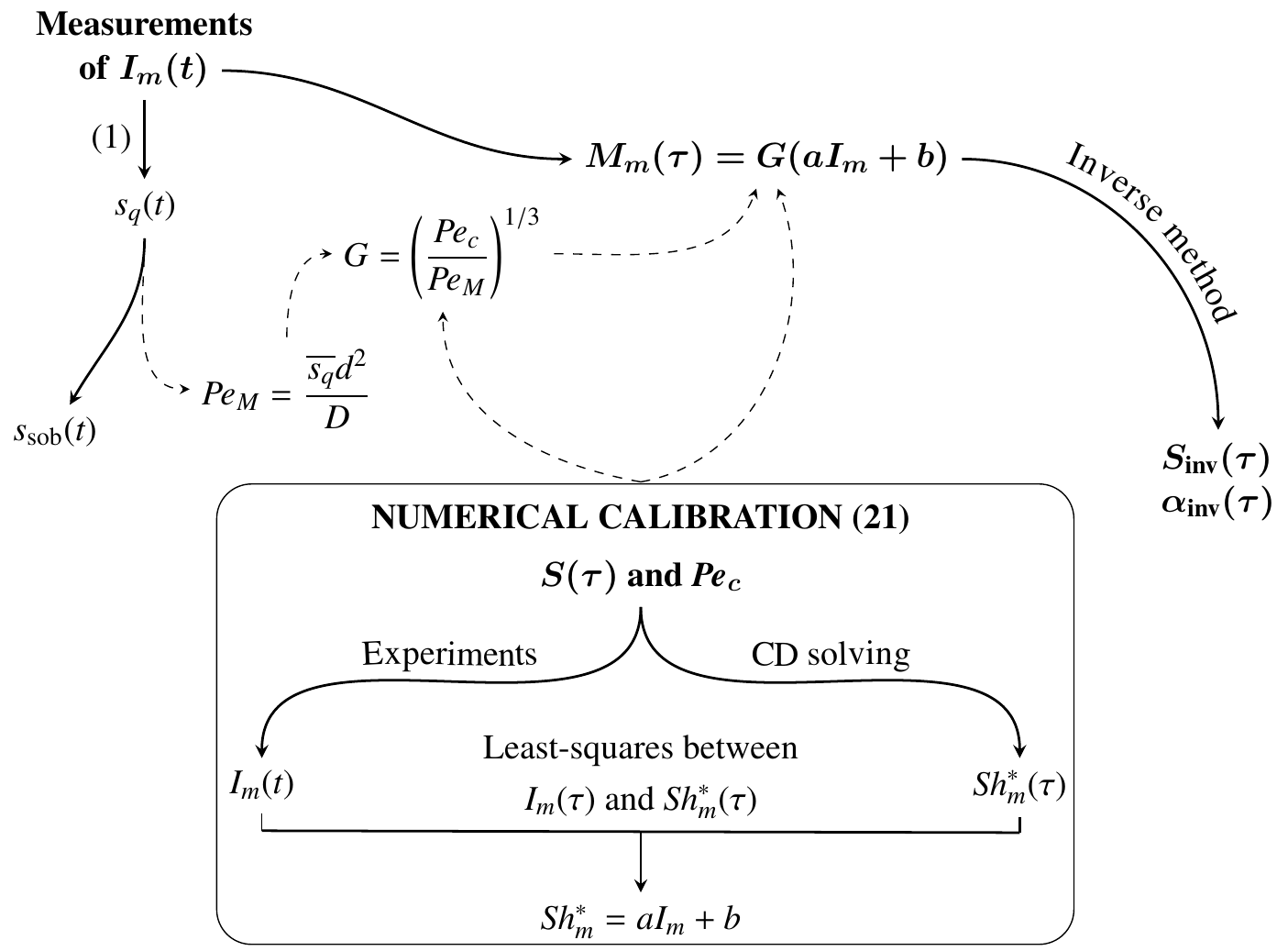}
	\caption{Proposed procedure for the nondimensionalization process.}
	\label{fig:17}
\end{figure*}

On top of the uncertainties discussed above, one should also consider the numerical modeling in \eqref{eq:Shdef2}. In fact, considering the existence of
\begin{enumerate}
	\item electrical noises (DC and/or AC, potentially altering the measured current with offsets, gains and random noise);
	\item parasite currents or side reactions caused by impurities in the solution;
	\item phenomena restricting the reaction (i.e. limiting current condition not achieved ($C_\text{probe}\neq0$), probe poisoning, lost of active area),
\end{enumerate}
one could suspect that the CD equation alone cannot take into account such complex effects. Even in a steady process, the ratio $\Sh/I$, evaluated using the current from real probes and the Sherwood number obtained in the corresponding steady direct problem (same $\Pen$, $S=1$), can sometimes vary by $\sim1-2\,\%$ per $\Pen$ decade as observed in Fig.~\ref{fig:15}, leading to errors on $\ol{s}$ up to 6\,\% as per \eqref{eq:Slev}. Such an error on $\ol{s}$ will alter the conversion $I\to M$ using \eqref{eq:shexp} and, accordingly, the resulting $S$ and $\alpha$ after the inverse process. The form $\she=KI$ suggested by \eqref{eq:Shdef2} is thus questionable, especially if unsteady evolutions characterized by a more sophisticated dynamic are also considered. An additional calibration is then proposed to cover the gap between measures and the numerical model, based on the form of \eqref{eq:Shdef2}:
%The , leading to $\ol{S}\neq1$; t
\begin{align}
	M =  G(\Pen)(aI+b), \label{eq:calibnum}
\end{align}
where the purpose of constants $a$ and $b$ is to adjust the measured current to match the numerical model and $G$ is a function to correct for the $\Sh^*$ variation with $\Pen$ owing to axial diffusion effects, defined as
\begin{align*}
	G(\Pen)	&= \Sh_\text{std}(\Pen_c)/\Sh_\text{std}(\Pen), \\
	&\simeq \left(\Pen_c/\Pen\right)^{1/3}, 
\end{align*}
with the $c$ and `std' subscripts referring to the calibration conditions and steady state values, respectively. The concept of this calibration is to match the measured current $I(t)$ from a quasi-steady process (for example, periodic oscillation of $s(t)$ at low-frequency and amplitude) with the equivalent $\Sh^*(\tau)$ from the direct problem resolution at the corresponding conditions (same $S(\tau)$, $\Pen$, $\Sr$); constants $a$ and $b$ are then calculated by executing a least-squares regression between $\Sh^*(\tau)$ and the function $F=aI+b$ to obtain the best fit between experimental data and the numerical model. The overall process is synthesized in Fig.~\ref{fig:17}.

\begin{figure*}
	\centering
	\includegraphics{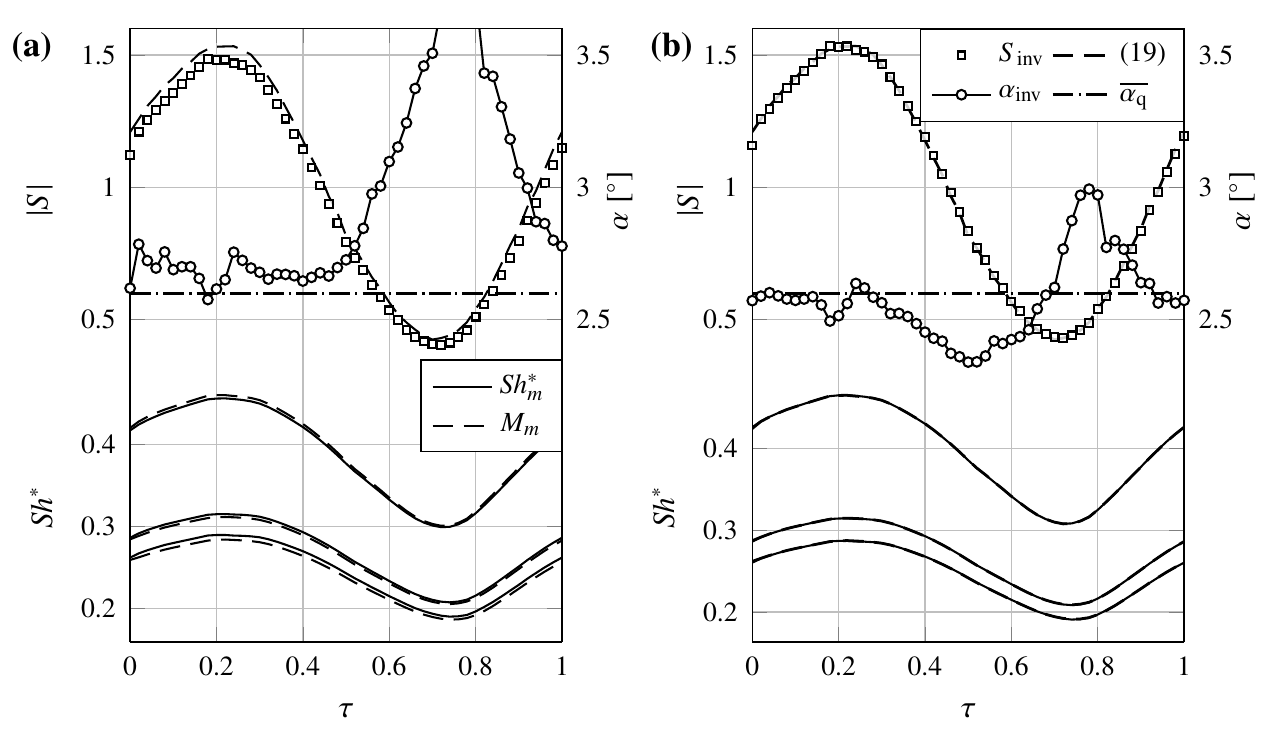}
	\caption{$S$, $\alpha$ and $\Sh^*_m$ obtained from the post-processing of a three-segment probe measurements with the inverse method at $\Pen=2.23\e{5}$, $\Sr=0.05$. Typical results when \textbf{a} direct conversion \eqref{eq:Shdef2} is used and \textbf{b} the form of \eqref{eq:calibnum} is opted for. At convergence, $\Sh^*_m$ are shifted from the experimental values in \textbf{a} while the convergence is perfect in \textbf{b}, affecting the precision on both $S$ and $\alpha$. One should note that at higher $\Sr$ and larger amplitude on $S$ (or $\alpha$), discrepancies between $M_m$ and $\Sh^*_m$ are likely to be amplified, especially when using \eqref{eq:Shdef2}. Are also present in top figures the analytical $S$ from \eqref{eq:Stot} along with the mean flow direction $\ol{\alpha_\text{q}}$; no reference value was available for $\alpha(\tau)$.}
	\label{fig:16}
\end{figure*}

Such a procedure is even more interesting considering the usage of a three-segment probe, where different coefficients can be determined for each segment. Hence, three $(a,b)$ couples are to be found, which will provide corrections for the geometrical discordances between each segment along with those between the real probe and the discretized one. 

The two-component wall shear rate obtained with the inverse method when using the proposed numerical calibration are plotted in Fig.~\ref{fig:16}b for a low-frequency, moderate-amplitude $S$ variation. One can notice that results so-obtained are flawless for $S$, while only a minor variation ($\sim \pm 0.5\,\deg$) characterizes the $\alpha(\tau)$ evolution, oscillating around the mean quasi-steady flow direction $\ol{\aq}$ retrieved from a directional calibration \citep{wein1987theory} (which is here considered as the \textit{true} direction considering the absence of shear reversal). Convergence of the inverse algorithm on the imposed fluctuations is fast and efficient, as a proper convergence for $S_\text{inv}$ is already noted after the very first time step (Fig.~\ref{fig:16}b). More time steps are although required at higher $\Sr$ (see Sect.~\ref{sec:results}). Without the use of the numerical calibration in the inverse process (i.e. when using \eqref{eq:Shdef2} instead), a proper convergence of all three $\Sh^*_m$ on the measured $M_m$ is likely to fail. Acceptable results can be expected, though being distorted and shifted as observed in Fig.~\ref{fig:16}a. Note that a similar \textit{steady} numerical calibration is also possible, using for instance data from Fig.~\ref{fig:15}. Least-squares could then be performed between $\Sh(\Pen)$ and $I(\Pen)$. This would, however, not take into account potential dissimilarities in unsteady flows.

Besides, recalling \eqref{eq:Slev}, a proper probe calibration gives access to the time-averaged wall shear rate without the need for electrochemical parameters such as $D$, $C_0$, $A$ and their associated errors, all regrouped in the constant $\klev$. While the same objective is intended for the numerical calibration \eqref{eq:calibnum}, $d$ and $D$ are at least needed for the indispensable calculation of $\Pen$ and $\Sr$. For large $\Pen$ and low $\Sr$ however, errors when evaluating those parameters only slightly affect the $\Sh^*$ values resulting from the direct or inverse problems considering their unimportant weight in the convection--diffusion and sensitivity equations; such flow conditions are thus recommended when performing the numerical calibration.

\section{Results and discussion \label{sec:results}}
While Fig.~\ref{fig:16} already demonstrated the effectiveness of the proposed procedure, advantages of using the inverse method are especially profitable when dealing with unsteady shear reversal or two-dimensional shear stress, as the combined \sob~and quasi-steady solutions fail to procure valid results \citep{lama2017inv}. However, only the former case will here be considered, as the latter is not achievable in the current experimental setup.

\begin{figure*}
	\centering
	\includegraphics{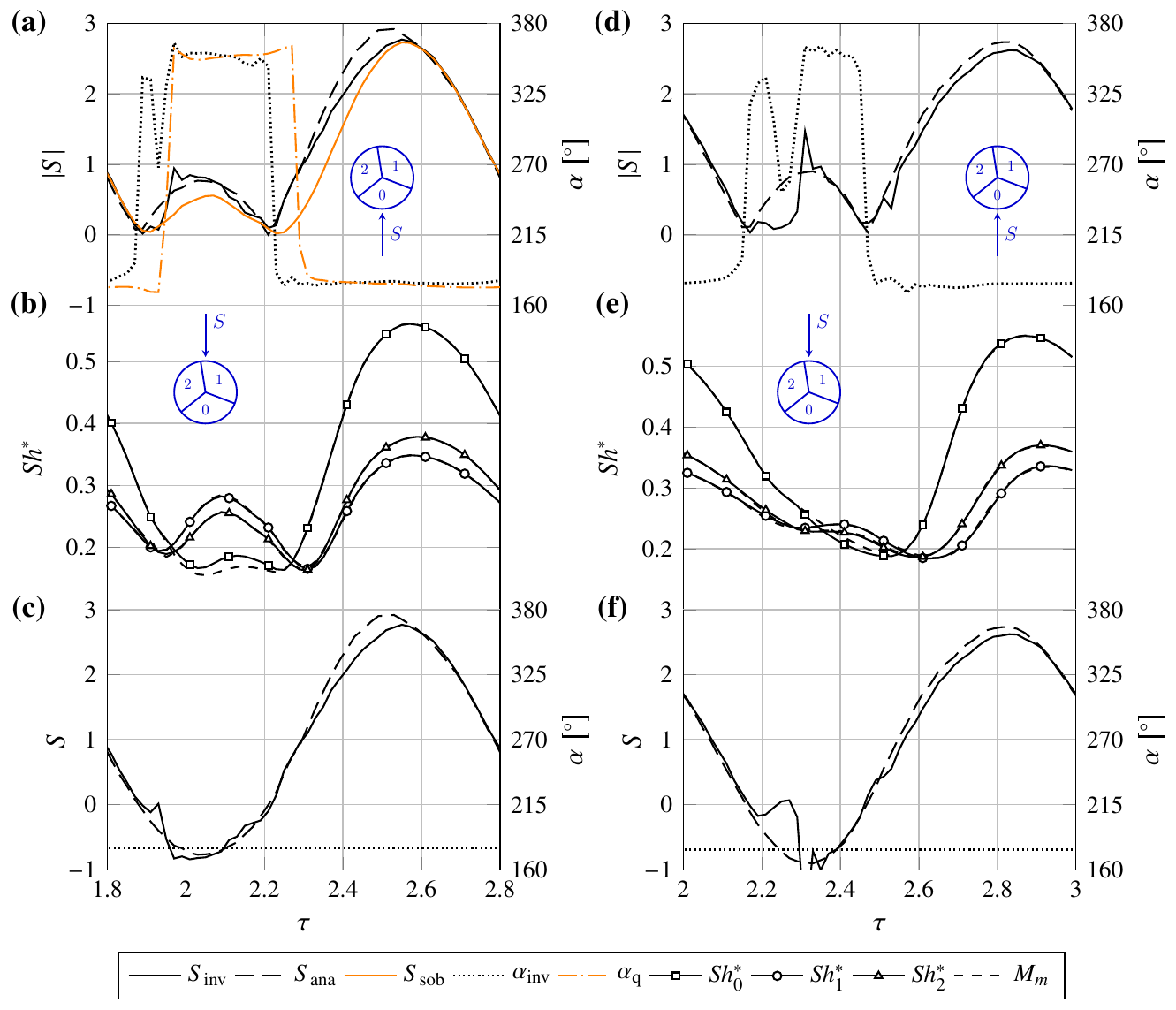}
	\caption{Results of the two-dimensional inverse method for \textbf{a, b} case 0 and \textbf{d, e} case 1. While $S$ can take negative value in the process itself, only its magnitude is plotted; 180$\deg$ is here added to $\alpha$ when $S<0$. A one-dimensional inverse algorithm was used to procure results shown in \textbf{c, f} for cases 0 and 1 respectively, where only the signed wall shear rate $S$ is solved for while  $\alpha$ is set to the average direction in the forward flow phase.}
	\label{fig:18}
\end{figure*}

Flow parameters for two cases involving shear reversal, which could be classified as low- and mid-frequency, are summarized in Table \ref{tab:parexp}. One should first note that while the actual frequencies $f$ associated with the present flows are rather low, this is not the case for the $\Sr$ number considering the size of the probe used (equivalent diameter $d\sim0.5\,$mm). Recalling \eqref{eq:cdadim3D}, $\Sr\sim1$ suggests that inertial effects are comparable to the diffusive ones in the $y$ direction. Using a probe five times smaller, the associated frequency for the same $\Sr$ as case 1 would be $\sim 100$\,Hz. Inverse method's results for both cases are shown in Fig.~\ref{fig:18} along with the analytical wall shear rate deduced from LDA measurements using the procedure of Sect.~\ref{sec:Sref}. While no reference values are available for $\alpha$, reversal phases in those one-dimensional wall shear rate flows are characterized by $\Sana<0$. 

\begin{table}[h]
	\caption{Flow parameters in the experiments.}
	\begin{tabular}{cccccc}
		\hline\noalign{\smallskip}
		Case & $f$\,[Hz]	& \,\,$\ol{s}$\,[s$^{-1}$]\,\, & $|s'_\text{max}|$\,[s$^{-1}$] & $\Pen$ 	  & $\Sr$ \\
		\noalign{\smallskip}\hline\noalign{\smallskip}
		0	 & 1.4		& 636				  & 1220			& $2.29\e{5}$ & 0.135 \\
		1	 & 4.0		& 528				  & 915				& $1.90\e{5}$ & 0.436\\
		\noalign{\smallskip}\hline
	\end{tabular}	
	\label{tab:parexp}
\end{table}

Results plotted in Fig.~\ref{fig:18} for both cases show interesting resemblances with the analytical values. In particular, behaviors of both $S$ and $\alpha$ are very well predicted in the lower frequency case of Fig.~\ref{fig:18}a. Steep variations of $\ainv$ accurately confine the shear reversal period compared to one predicted by the quasi-steady solution $\aq$, which is considerably out of phase even in the low-frequency case 0; such an effect was also noted in the numerical results of \citet{lama2017inv}. $S$ obtained with the \sob method, not valid in reversing flows, also exhibits larger deviations from $\Sana$. Although not shown in Fig.~\ref{fig:18}d, error associated with $\aq$ and $\Ssob$ are more important at higher $\Sr$. Main discrepancies regarding the inverse method results arise when the $\Sh^*_m$ curves cross each other. This first occurs in case 0 between segments 1 and 2 at $\tau\simeq1.92$ (Fig.~\ref{fig:18}b), merely after the inception of shear reversal, and a little later involving segments 0--1, then 0--2. During this transition, the $\Sh^*_m$ are of the same order of magnitude which, combined with the lower segment sensitivities at those low convection phases ($S\sim0$), blurs the inverse process for a short period; this is then followed by an unstable recovery marked with sharp spikes on both $S$ and $\alpha$, result of the lack of convergence as observed with the $\Sh^*_m$ evolutions. After such discontinuities, a few time steps are often needed to stabilize and converge back on proper values for $S$--$\alpha$, here well exposed with the $\Sh^*_0$ curve. The following $\Sh^*_m$ intersections, occurring after the shear reversal period, still alter the progression. Results shown in Fig.~\ref{fig:18} were obtained after 2--3 periods of calculation after which no further convergence improvements are observed. As these episodes happen at the same phases periodically, the inverse process can never completely converge on the exact $\Sh^*_m$ evolutions and hence on the `true' $S$--$\alpha$. In the higher frequency case (Figs.~\ref{fig:18}d--e), errors are more apparent in the shear reversal period, which could be explained by the lower sensitivities at higher $\Sr$ \citep{lama2017inv}. Furthermore, one can observe that the $\Sh^*_1$ and $\Sh^*_2$ curves are flattened and merely superposed during the reversal phase, adding to the blurring effect. Despite those convergence issues, overall evolutions are well predicted and it is interesting to note that, from such complex and phase shifted signals as the $I_m(t)$ curves shown in Fig.~\ref{fig:4}, one is able to retrieve a proper two-component instantaneous wall shear rate.

Figures \ref{fig:18}c, f show results of the inverse process when $\alpha$ is not solved for and a constant direction is imposed, which here corresponds to the average direction in the forward flow phase. This undoubtedly simplifies the inverse problem, the only unknown being the signed wall shear rate $S$. Results are very similar to those of Figs.~\ref{fig:18}a, d for $S$, without any significant improvement nor deterioration besides the enhanced convergence speed as less iterations are required in the inverse problem. However, one should notice that such a procedure is different than the one-dimensional inverse problem as performed by \citet{mao1991analysis} for the circular probe, since the flow direction is not imposed. All three signals are still used in the inverse process, hence improving the shear reversal detection. Obviously, this is only consistent for one-dimensional wall shear rate flows. One should further note that although the $S$ and $\alpha$ evolutions of cases 0 and 1 can be considered one-dimensional, results with the present two-dimensional inverse problem demonstrates that the proposed algorithm can deal with very steep variations of the variables like those observed for $\alpha(\tau)$ in Figs.~\ref{fig:18}a, d. The fact that these sharp fluctuations occur when $S\sim0$, i.e. when the probe sensitivity is the lowest, might also explain the poorer convergence in the shear reversal period. Investigations in flows exhibiting a more complex $\alpha(\tau)$ variation while featuring stronger convection effects should be performed to improve the method and complete the validation process.

\begin{figure}
	\centering
	\includegraphics[]{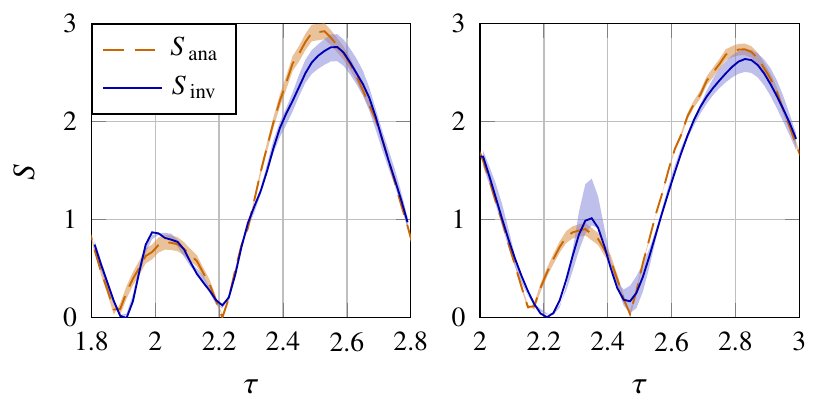}
	\caption{Errors associated with $\Sana$ and $\Sinv$, illustrated with filled areas for case 0 (left) and case 1 (right). Errors on $\Sana$ is calculated using $h\pm0.1\,$mm in \eqref{eq:Stot} while that on $\Sinv$ suppose a combined error of $\pm5\,\%$ on the parameter $d^2/D$, which is founded in both $\Pen$ and $\Sr$ definitions. Note that results were filtered for visualization purposes.}
	\label{fig:13}
\end{figure}

\subsection{Notes on discrepancies and measurement errors \label{sec:measerr}}
Two main issues are essentially exposed in the results of Fig.~\ref{fig:18}. Besides the lack of a proper convergence in the shear reversal period, one can also note that the maximal $S$ value in both cases 0 and 1 is lower than the analytical one. While uncertainties associated with both ED and LDA methods along with those related to the analytical solution \eqref{eq:Stot} might be enough explain those discrepancies (cf. Fig~\ref{fig:13}), a few potential sources of error will be inspected in the following.

\begin{figure*}
	\centering
	\includegraphics{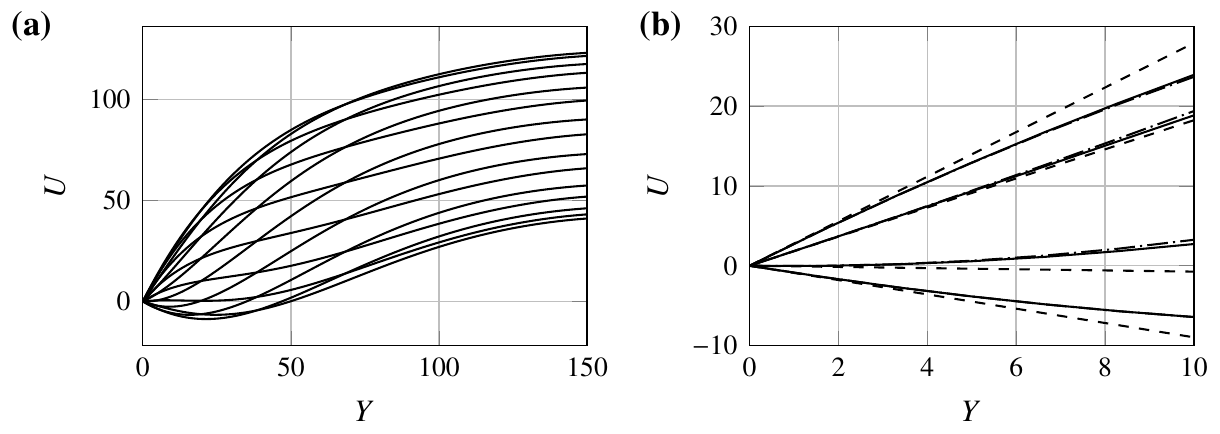}
	\caption{Analytical solution \eqref{eq:ucomplet} for case 1. Velocity is nondimensionalized using \eqref{eq:adim} and so $U=u\Pen^{1/3}d^{-1}\ol{s}^{-1}$. \textbf{a} Velocity profiles in the channel's lower half at different phases. \textbf{b} Close up in the region $Y<10$ for certain phases. Dashed lines represent the classical linear velocity profile $U=SY$; dashed--dotted ones are for the quadratic approximation \eqref{eq:uquad}.}
	\label{fig:19}
\end{figure*}

\begin{figure*}
	\centering
	\includegraphics{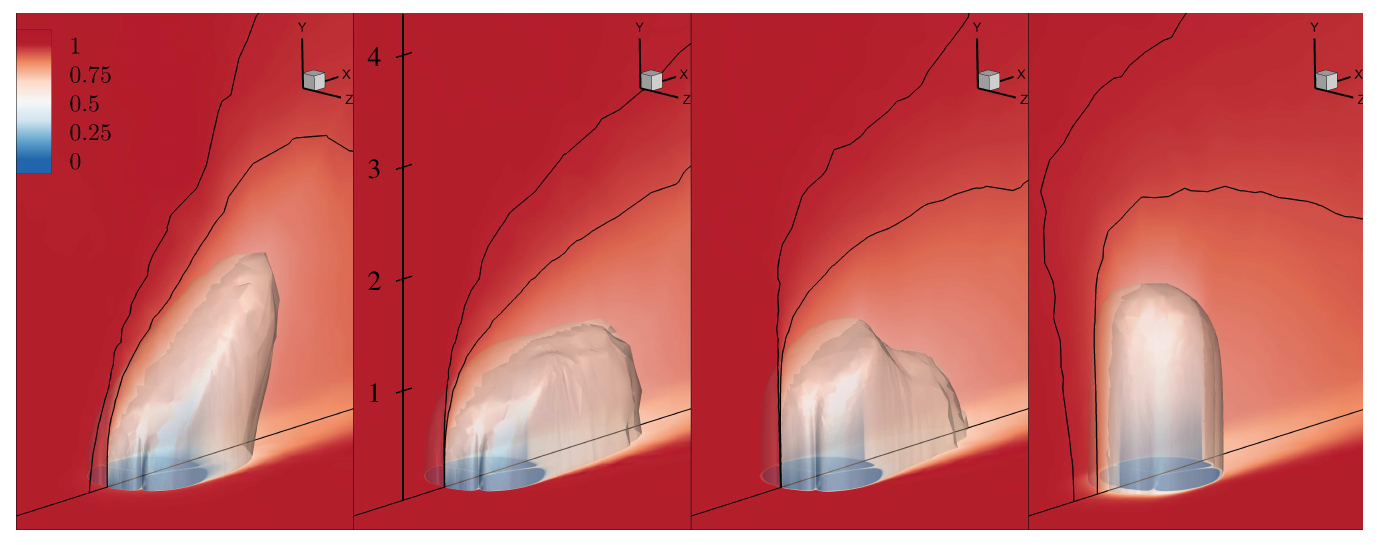}
	\caption{Concentration field associated with flow conditions of case 1 at four successive time steps. From left to right: $\tau=\{0.22,0.44,0.62,0.98\}$. Black contour lines indicate $C=\{0.9,0.99\}$. Also shown are the concentration iso-surfaces for $C=0.7$.}
	\label{fig:20}
\end{figure*}

\begin{figure}[h!]
	\centering
	\includegraphics[]{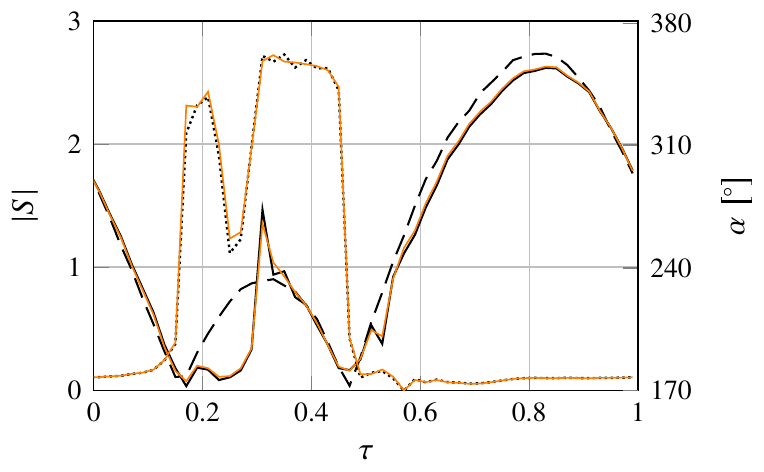}
	\caption{Effect of using a quadratic velocity profile (colored curves) in \eqref{eq:cdadim3D} compared to the linear one (black curves, same data and line patterns as in Fig.~\ref{fig:18}a, case 1).}
	\label{fig:12}
\end{figure}

One important hypothesis of the inverse method is the imposition of a linear velocity profile in the convection--diffusion equation. As observed in Fig.~\ref{fig:19}b, considerable differences are noticed between the analytical velocity profile and a linear one at certain time steps. Although these are observed far from the major concentration gradients, located at $Y\lesssim1$ in the steady-state diffusion layer, its thickness $\delta$ can grow much higher in reversal periods depending on the solicitation amplitude and frequency. Regarding case 1, the $C=0.99$ contour line extracted from the numerical simulation can go up to $Y=4$ above the probe (Fig.~\ref{fig:20}). In this highly convective region, Fig.~\ref{fig:19} exposes larger discrepancies with the analytical solution; one could then expect the linear velocity profile to alter the `true' diffusion layer and hence the probe's response. A quadratic velocity profile constructed with the normal derivative of the analytical wall shear rate equation \eqref{eq:Sn},
\begin{align}
\pd{s}{y} &= \left.\pdd{u}{y}\right|_{y=\pm h/2} \nonumber \\
&= \sum_{n=1}^{N} \frac{iK_n}{\nu}\me^{i\omega_nt}, 
\label{eq:dSn}   
\end{align}
is also shown in Fig.~\ref{fig:19}b, offering a fairly more accurate approximation of the analytical solution in the region $Y<10$. Such a velocity profile has then been employed in the direct problem equation \eqref{eq:cdadim3D}, where the velocity $U$ is replaced by its second order Taylor development at the wall, namely
\begin{align}
U = SY + \frac{Y^2}{2}\pd{S}{Y},
\label{eq:uquad}
\end{align}
where, from \eqref{eq:adim}, \eqref{eq:u0} and \eqref{eq:dSn}, 
\begin{align}
\pd{S}{Y} = \left(-\frac{2\ol{s}}{h} + \sum_{n=1}^{N} \frac{iK_n}{\nu}\me^{i\omega_nt} \right) \frac{d}{\ol{s}\Pen^{1/3}}.
\label{eq:dSdY}
\end{align}
While $\partial S/\partial Y$ cannot be directly assessed from ED signals, it is interesting to note from \eqref{eq:dSdY} that it can be deduced from LDA measurements using the same $K_n$ as the one calculated in \eqref{eq:Sn}. Alternatively, if one had access to instantaneous wall pressure measurements, no-slip condition allows to rewrite \eqref{eq:momen} so as to obtain the same information, that is
\begin{align}
\pd{s}{y} = \frac{1}{\mu}\pd{p}{x},
\end{align}
although this has not been tested experimentally. Hence, using \eqref{eq:dSdY} as additional information on the velocity profile curvature, one could then investigate its benefits on the inverse method results, if any. Note that the only change is the added term in the direct problem, while the inverse problem and algorithm stay unaltered. A quadratic velocity profile, where the curvature is known and imposed at each time step, was thus used in the post-processing of both cases 0 and 1. It was found (Fig.~\ref{fig:12}) that although results so-obtained are closer to the analytical values, the amplitude of $S$ only slightly increases and still does not match the analytical values, nor is the $\Sh^*_m$ convergence improved in the shear reversal period except from a somewhat smoother $\alpha$ evolution. Such interesting results, while unable to justify the aforementioned discrepancies, reinforce the validity of the ED fundaments and the method's application range; the linear velocity profile approximation in \eqref{eq:cdadim3D} thus remains adequate even when strong curvature characterizes the close-wall velocity profile. 

\begin{table*}
	\centering
	\caption{Effect of the $\ol{s}$ value used in the nondimensionalization relation \eqref{eq:adim}. When using the overestimated $\ol{s_q}$ in the shear reversing cases 0--1, $S$ obtained in the inverse process is down-shifted ($\ol{S}<1$). $\ol{s}_\text{ana}$, deduced from LDA measurements, can here be considered as the true value despite the 0.95 value obtained in case 0, which is attributed to inverse convergence issues and measurement errors discussed in Sect.~\ref{sec:measerr}. Actual values are $\ol{s}_\text{ana}/\ol{s}_q\simeq0.92$ for both cases 0 and 1 as per probe calibration.}
	\begin{tabular}{c}
		\hline\noalign{\smallskip}
		\begin{tabular}{ccc}
			&  \multicolumn{2}{c}{$\ol{S}$ after inverse process} \\ 
			$\quad$ Value used $\quad$  & \multirow{2}{*}{$\quad$Case 0$\quad$}  
			& \multirow{2}{*}{$\quad$Case 1$\quad$} \\
			for $\ol{s}$   		&  &  \\ \hline 
			$\ol{s}_q$   		& 0.87    & 0.91     \\
			$\ol{s}_\text{ana}$ & 0.95    & 0.99                     
		\end{tabular}
		\hspace{2.em}
		\begin{tabular}{ccc}
			&  \multicolumn{2}{c}{Actual $\ol{s}$ values [s$^{-1}$]} \\ 
			& \multirow{2}{*}{$\quad$Case 0$\quad$}  
			& \multirow{2}{*}{$\quad$Case 1$\quad$} \\
			&  &  \\ \hline
			$\ol{s}_q$   		& 696    & 575     \\
			$\ol{s}_\text{ana}$ & 636    & 528                        
		\end{tabular}\\
		\noalign{\smallskip}\hline
	\end{tabular}

	\label{tab:smean}
\end{table*}

As further verification, raw three-segment probe signals (see Fig.~\ref{fig:4}a) were used in the inverse process instead of the phase-averaged one. Apart from additional noise in the results, similar evolutions as those observed in Fig.~\ref{fig:18} were obtained. Latest potential cause arises from the actual $I\to M$ conversion as per \eqref{eq:calibnum}. Considering that the ratio $\Sh/I$ can vary with $\Pen$ as observed in Fig.~\ref{fig:15}, the linear form of \eqref{eq:calibnum} may actually be too elementary when both shear rate amplitude and frequency are large. Indeed, during phases when $S(\tau)>1$, the local convection gets two to three times stronger than the average $\ol{S}=1$ state (cf. Fig.~\ref{fig:18}). One might then seek a form based on an unsteady $\Pen$ number which would consider the local flow acceleration at any time $\tau$. The form of \eqref{eq:calibnum} or the overall calibration process may thus need to be revisited for further improvements. Moreover, considering that migration effects are not completely suppressed and that the Nernst diffusion layer\footnote{At the electrode surface, a stagnant diffusion layer $\delta$ is assumed \citep{bard2001electrochemical}, hypothesis at the basis of \eqref{eq:shexp} and \eqref{eq:shnum}.} approximation in unsteady flows is questionable, equations \eqref{eq:shexp} and \eqref{eq:shnum} used to evaluate $\Sh^*$ might be to reconsider. Hence, the $I\to M$ conversion may be the main cause regarding the light attenuation of $S(\tau)$. Nonetheless, recalling the discussion of Sect.~\ref{sec:calnum} and the excellent results when $S>0$ (cf. Fig.~\ref{fig:16}), the proposed procedure definitely helps to reduce impacts of measurement errors. 

On the other hand, convergence issues in the shear reversal period could also be explained by a lack of probe sensitivity due to the size of the interstices. Poorer convergence properties were indeed confirmed by numerical analysis when modeling the sensor with larger gaps \citep[see also][]{lama2017inv}. Furthermore, in the numerical tests conducted by \citet{lama2017inv}, the shear reversal was accurately predicted throughout the cycle in a similar reversing flow. Additional efforts should thus be committed to reduce the gap size of real three-segment probes and improve the experimental technique. The inverse algorithm itself should also be further investigated. 

One should lastly note that in flows exhibiting shear reversal like cases 0 and 1, the use of the quasi-steady solution \eqref{eq:Slev} for $\ol{s}$ evaluation leads to erroneous values for $\Pen$ and $\Sr$. As a result, condition $\ol{S}=1$ is not fulfilled after the inverse process, shifting $S(\tau)$ toward the value $\ol{S}=\ol{s}/\ol{s_q}$, where $\ol{s}$ is the \textit{true} time-averaged wall shear rate (see Table \ref{tab:smean}). A complementary method is then preferred for the measure of $\ol{s}$; otherwise, one could iteratively correct the average value after each period of the inverse process until $\ol{S}=1$ is obtained, considering a proper numerical calibration was performed. Then only, one can expect that the corrected $\ol{s}$ value is appropriate following a converged inverse process. As such a procedure was irrelevant in the present study, $\ol{s}$ values were directly taken from LDA measurements.

%In the authors opinion, one of the main source of errors would reside in the evaluation of $S(\tau)$ in the calibration process (see Fig.~\ref{fig:17}) which will affect parameters obtained in the least square fitting as well as the $G$ function. As discussed in \S\ref{sec:Sref}, the mean wall shear stress error was evaluated to $\sim5\%$ in the present experimental apparatus in regards of uncertainties on $h$. This error also affect the unsteady part as per \eqref{eq:Sn}. Alternative calibrations were performed using $h\pm 0.1~$mm in $\eqref{eq:Stot}$, \note{but $S$ resulting from the inverse process was not improved nor was the discrepancies with the analytical solution diminished}. 

\section{Concluding remarks}
The two-component wall shear rate under a high-amplitude pulsed channel flow was assessed with mass transfer probes. When post-processing the signals using the two-dimensional inverse method \citep{lama2017inv}, the resulting wall shear magnitude and direction showed strong concordance with the known instantaneous values, available from LDA measurements. As the solicitation frequency was increased, the inverse problem presented some convergence issues in part of the shear reversal phase, leading to small discrepancies with the analytical solution. Some potential causes were examined. It was shown that the use of a qua\-dra\-tic velocity profile in such flows exposing shear reversal did not bring substantial improvements in the results, reinforcing the idea that the linear $U=SY$ profile should be sufficient in most applications. Fabrication of a sensor with smaller gaps should improve its sensitivity to shear reversal and facilitate the inverse process. A nondimensionalization procedure was also proposed when treating experimental data with the inverse method. This approach allows one to bond the numerical model to the experimental one, both in terms of the sensor discretization and approximations in the model itself, which for instance cannot take into account potential side reactions that could occur during the experiments. Besides, when dealing with a high-amplitude wall shear rate fluctuation in a non-reversing flow, the inverse method using the proposed calibration procedure offered a precise wall shear rate magnitude evolution and accurate direction. The present work thus exposes the potential of ED probes to deal with complex three-dimensional unsteady flows when coupled with the proposed method. Experiments involving a time varying direction should be performed for further verifications and improvements of the two-dimensional inverse problem, even though the method was numerically validated for such a case \citep{lama2017inv}.

\begin{acknowledgements}
\sloppy
The authors would like to acknowledge the financial support of the Canadian Foundation for Innovation (CFI), the Natural Sciences and Engineering Research Council of Canada (NSERC) and the Fonds de recherche du Qu\'ebec - Nature et technologies (FRQNT). We also acknowledge the technical support of T. Lafrance from M\"E\-KA\-NIC and J.-M. B\'eland in the design and manufacturing of the experimental setup.
\end{acknowledgements}

\end{document}